\documentclass[11pt,twoside]{article}
\usepackage{asp2010}

\resetcounters

\newcommand{\lSect}[1]{{\label{sec:#1}}}
\newcommand{\lFig}[1]{{\label{fig:#1}}}

\def\gtaprx {\lower .1ex\hbox{\rlap{\raise .6ex\hbox{\hskip .3ex
    {\ifmmode{\scriptscriptstyle >}\else
        {$\scriptscriptstyle >$}\fi}}}
    \kern -.4ex{\ifmmode{\scriptscriptstyle \sim}\else
        {$\scriptscriptstyle\sim$}\fi}}}
\def\ltaprx {\lower .1ex\hbox{\rlap{\raise .6ex\hbox{\hskip .3ex
    {\ifmmode{\scriptscriptstyle <}\else
        {$\scriptscriptstyle <$}\fi}}}
    \kern -.4ex{\ifmmode{\scriptscriptstyle \sim}\else
        {$\scriptscriptstyle\sim$}\fi}}}
\newcommand{\note}[1]{\emph{\textcolor{red}{}}}

\newcommand{\Msun}{{\ensuremath{\mathrm{M}_{\odot}}}}

\newcommand{\Ni}{{\ensuremath{^{56}\mathrm{Ni}}}}

\newcommand{\FIGFF}[2]{{\ref{fig:#2}{#1}}}

\newcommand{\FIG}[2]{{Fig.~\FIGFF{#1}{#2}}}
\newcommand{\Fig}[1]{{\FIG{}{#1}}}

\newcommand{\CASTRO}{\texttt{CASTRO}}
\newcommand{\KEPLER}{\texttt{KEPLER}}
\begin{document}

\title{Multidimensional Simulations of Thermonuclear Supernovae from the First Stars}
\author{ Ke-Jung Chen$^1$,  Alexander Heger$^1$, and Ann Almgren$^2$
\affil{$^1$ School of Physics and Astronomy, University of  Minnesota, Minneapolis, MN 55455, USA}
\affil{$^2$ Center for Computational Sciences and Engineering, Lawrence Berkeley National Laboratory, Berkeley, CA 94720, USA}}

\begin{abstract}

Theoretical models suggest that the first stars in the universe could have been very massive, with typical 
masses $\gtrsim$ 100 \Msun. Many of them might have died as energetic thermonuclear 
explosions known as pair-instability supernovae (PSNe).  We present multidimensional 
numerical simulations of PSNe with the new radiation-hydrodynamics code CASTRO.  
Our models capture all explosive burning and follow the explosion until the shock breaks 
out from the stellar surface. We find that fluid instabilities driven by oxygen and helium 
burning arise at the upper and lower boundaries of the oxygen shell $\sim$ 20 - 100 sec 
after the explosion begins.  Later, when the shock reaches the hydrogen envelope a 
strong reverse shock forms that rapidly develops additional Rayleigh-Taylor instabilities. 
In red supergiant progenitors, the amplitudes of these instabilities are sufficient to mix the 
supernova's ejecta and alter its observational signature. Our results provide useful 
predictions for the detection of PSNe by forthcoming telescopes.

\end{abstract}

\section{Introduction}

The evolution of the first stars in the universe is one of the frontiers of modern cosmology. 
Primordial stars synthesized the first heavy elements in the universe, and their energetic 
feedback influenced the formation of later generations of stars and the first galaxies \citep{
wet08a,wet08b,greif2010}.  Early numerical models predicted that Pop III stars formed with 
masses of 100-1000 \Msun{} \citep{bromm2009,abel2002}.  New studies have found that 
$\sim$ 20\% of Pop III stars form in binaries or multiples \citep{turk2009,stacy2010} so the 
first stars could be less massive than originally thought.  However, even today observations 
support the existence of stars with initial masses over 150 \Msun\  \citep{r136}.  Stellar 
evolution models predict that Pop III stars with initial masses of 140 - 260 \Msun{} develop 
oxygen cores of $\gtrsim$ 50 
\Msun{} after central carbon burning \citep{heger2002}. At this point the core reaches sufficiently 
high temperatures ($\sim$ $10^9$ K) and low densities ($\sim$ $10^6$ g/cc) that the creation 
of electron-positron pairs is favored. Radiation pressure support then quickly decreases, triggering 
a rapid contraction of the core.  During contraction, core temperatures and densities sharply rise 
and oxygen and silicon begin to burn explosively. The resulting thermonuclear explosion, known 
as a pair-instability supernova (PSN), reverses the contraction and completely unbinds the star, 
leaving no compact remnant and forming up to 50 \Msun\ of \Ni{}. One possible PSN candidate, 
SN 2007bi, has recently found by \citep{2007bi}.  
  
Most current theoretical models of PSNe are based on one-dimensional calculations \citep{
heger2002}. However, in the initial stages of a supernova spherical symmetry is 
broken by fluid instabilities generated by burning, which cannot be captured in 1D.  
Two-dimensional simulations of Pop III PSNe have recently been done by \citet{candace2010-2}
in which only mild dynamical instabilities were found to form, but they proceeded from 1D KEPLER 
models in which explosive burning had already occurred and thus exclude instabilities driven by 
burning.  Such instabilities, if they form, may alter the energetics and nucleosynthesis of the SN 
by vigorously mixing its fuel and must be included in simulations to understand the true evolution 
of PSNe.  We have performed 2D simulations of Pop III PSNe that follow the initial contraction of the 
core until most of the energy due to explosive burning has been released, in contrast to
\citet{candace2010-2}, who only follow the post-nucleosynthesis hydrodynamics.  Our goal is to study any 
fluid instabilities that arise and how mixing alters nucleosynthesis and the energetics of the explosion.  

\section{Numerical Methods}

\lSect{method}

We evolve zero-metallicity stars in \KEPLER{} \citep{kepler}, a one-dimensional Lagrangian 
stellar evolution code.  In \KEPLER{} we solve evolution equations for mass, momentum, and 
energy and include physics relevant to stellar evolution such as nuclear burning and artificial 
mixing. When the star comes to the end of central oxygen burning, we map its profile onto a 
2D Cartesian grid in \CASTRO{}. The procedure for mapping and seeding initial perturbations 
in these profiles is discussed in detail in \citet{chen2011-2}. We evolve the star in \CASTRO{} 
through the end of explosive burning. 

\CASTRO{} \citep{ann2010,zhang2011} is a massively parallel, multidimensional Eulerian 
adaptive mesh refinement (AMR) radiation-hydrodynamics code for astrophysical applications. 
Its time integration of the hydrodynamics equations is based on a higher-order, unsplit Godunov 
scheme. Block-structured AMR with subcycling in time enables the use of high spatial resolution 
where it is most needed. We use the Helmholtz equation of state (EOS) \citep{timmes2000} with 
density, temperature, and species mass fractions as inputs. The gravitational field is 
calculated using a monopole approximation constructed from a radial average of the 2D density
field on the grid.

\begin{figure}
  \begin{center}
    \includegraphics[width=6 cm]{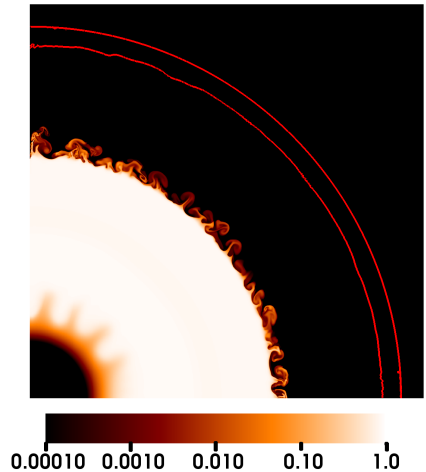}
  \end{center}
  \caption{Fluid instabilities at early stages of the explosion.  We show oxygen mass fraction 100 
  sec after the reversal of core collapse. Dynamical instabilities appear at the lower boundary of 
  the O shell because of oxygen burning and at the upper boundary of the shell because of He
  burning. The red lines indicate the location of shock.   \lFig{PSN-1}}
\end{figure}

\section{2D Simulations} 

\lSect{results}

In \Fig{PSN-1} we show the formation of dynamical instabilities at the base of the oxygen burning 
shell during the contraction of the core \citep{chen2011-3}.  They are relatively mild and do not 
penetrate the central \Ni\ region, so no \Ni\ is mixed into the upper layers of the star at this stage.  
After explosive burning reverses the contraction of the core, fluid instabilities driven by helium 
burning also appear in the outer layers of the oxygen shell. Minor mixing caused by these 
instabilities begins about 100 sec after reversal of the collapse.

As we show in \Fig{PSN-2}, when the shock propagates into the hydrogen envelope the 
formation of a strong reverse shock creates additional Rayleigh-Taylor instabilities (RTI).  Their 
amplitudes are sufficient to mix oxygen with the surrounding shells: H, He, and Si.  Some mixing 
also occurs at the outer edge of the \Ni{} core.  Our results demonstrate that dynamical instabilities 
form at several stages of the explosion and that they are mainly driven by RTI at the interfaces of 
contact discontinuities in density or species abundance. Mixing is important to the observational 
signatures of Pop III PSNe because it can cause absorption lines of heavy elements to appear in
spectra sooner after shock breakout than when mixing is absent.  The instabilities can also lead 
to the formation of clumps in the ejecta that can strongly affect its luminosity at later times. 

\begin{figure}
  \begin{center}
    \includegraphics[width=6 cm]{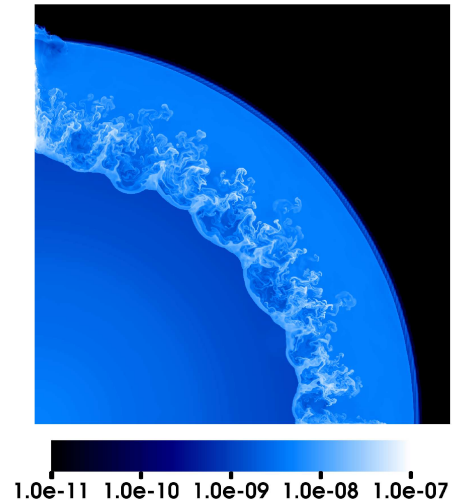}
  \end{center}
  \caption{Fluid instabilities prior to shock breakout.  Here, we show the density when the shock is 
  about to break out from the stellar surface. The fluid instabilities are driven by the reverse shock 
  and lead to significant mixing. \lFig{PSN-2}}
\end{figure}

\section{Conclusions}
\lSect{con}

In contrast to earlier multidimensional simulations, we find that mixing can occur at several stages 
in Pop III PSNe prior to shock breakout when core contraction and explosive nuclear burning is 
followed in 2D.  Mixing appears in PSN spectra by introducing absorption lines of heavy elements 
at early times and by changing its luminosity.  Our 2D simulations are the first in a numerical campaign 
to investigate the evolution of PSNe from their earliest stages as pre-supernova progenitors.  We are 
currently preparing to post-process our simulations to compute light curves and spectra in order to
provide useful predictions for the James Webb Space Telescope, which may soon detect these
primordial explosions.

\acknowledgements 
The authors would like to thank Daniel Whalen for reviewing the earlier manuscript and providing 
many insightful comments. We thank members of the CCSE at LBNL for their assistance with CASTRO. 
We also thank John Bell, Adam Burrows, Volker Bromm, and Stan Woosley for many useful discussions. 
This project has been supported by the DOE SciDAC program under grants DOE-FC02-01ER41176, DOE-FC02-06ER41438, and DE-FC02-09ER41618.   Finally, we acknowledge support from the AAS and NSF in the form of an International Travel Grant and from the 2011 Cefalu Meeting SOC/LOC in the form of the Best Abstract Prize, which enabled KC to attend this conference.

\bibliographystyle{asp2010}
\bibliography{Chen_K}

\begin{thebibliography}{}
\expandafter\ifx\csname natexlab\endcsname\relax\def\natexlab#1{#1}\fi
\expandafter\ifx\csname url\endcsname\relax
  \def\url#1{\texttt{#1}}\fi
\expandafter\ifx\csname urlprefix\endcsname\relax\def\urlprefix{URL }\fi
\providecommand{\eprint}[2][]{\url{#2}}

\bibitem[{{Abel} et~al.(2002){Abel}, {Bryan}, \& {Norman}}]{abel2002}
{Abel}, T., {Bryan}, G.~L., \& {Norman}, M.~L. 2002, Science, 295, 93.
  \eprint{arXiv:astro-ph/0112088}

\bibitem[{{Almgren} et~al.(2010){Almgren}, {Beckner}, {Bell}, {Day}, {Howell},
  {Joggerst}, {Lijewski}, {Nonaka}, {Singer}, \& {Zingale}}]{ann2010}
{Almgren}, A.~S., {Beckner}, V.~E., {Bell}, J.~B., {Day}, M.~S., {Howell},
  L.~H., {Joggerst}, C.~C., {Lijewski}, M.~J., {Nonaka}, A., {Singer}, M., \&
  {Zingale}, M. 2010, \apj, 715, 1221. \eprint{1005.0114}

\bibitem[{{Bromm} et~al.(2009){Bromm}, {Yoshida}, {Hernquist}, \&
  {McKee}}]{bromm2009}
{Bromm}, V., {Yoshida}, N., {Hernquist}, L., \& {McKee}, C.~F. 2009, \nat, 459,
  49. \eprint{0905.0929}

\bibitem[{{Chen} et~al.(2011{\natexlab{a}}){Chen}, {Heger}, \&
  {Almgren}}]{chen2011-2}
{Chen}, K., {Heger}, A., \& {Almgren}, A.~S. 2011{\natexlab{a}}, in preparation

\bibitem[{{Chen} et~al.(2011{\natexlab{b}}){Chen}, {Heger}, \&
  {Almgren}}]{chen2011-3}
--- 2011{\natexlab{b}}, in preparation

\bibitem[{{Crowther} et~al.(2010){Crowther}, {Schnurr}, {Hirschi}, {Yusof},
  {Parker}, {Goodwin}, \& {Kassim}}]{r136}
{Crowther}, P.~A., {Schnurr}, O., {Hirschi}, R., {Yusof}, N., {Parker}, R.~J.,
  {Goodwin}, S.~P., \& {Kassim}, H.~A. 2010, \mnras, 408, 731.
  \eprint{1007.3284}

\bibitem[{{Gal-Yam} et~al.(2009){Gal-Yam}, {Mazzali}, {Ofek}, {Nugent},
  {Kulkarni}, {Kasliwal}, {Quimby}, {Filippenko}, {Cenko}, {Chornock}, {Drake},
  {Thomas}, {Bloom}, {Poznanski}, {Miller}, {Foley}, {Silverman}, {Arcavi},
  {Ellis}, \& {Deng}}]{2007bi}
{Gal-Yam}, A., {Mazzali}, P., {Ofek}, E.~O., {Nugent}, P.~E., {Kulkarni},
  S.~R., {Kasliwal}, M.~M., {Quimby}, R.~M., {Filippenko}, A.~V., {Cenko},
  S.~B., {Chornock}, R., {Drake}, A.~J., {Thomas}, R.~C., {Bloom}, J.~S.,
  {Poznanski}, D., {Miller}, A.~A., {Foley}, R.~J., {Silverman}, J.~M.,
  {Arcavi}, I., {Ellis}, R.~S., \& {Deng}, J. 2009, \nat, 462, 624.
  \eprint{1001.1156}

\bibitem[{{Greif} et~al.(2010){Greif}, {Glover}, {Bromm}, \&
  {Klessen}}]{greif2010}
{Greif}, T.~H., {Glover}, S.~C.~O., {Bromm}, V., \& {Klessen}, R.~S. 2010,
  \apj, 716, 510. \eprint{1003.0472}

\bibitem[{{Heger} \& {Woosley}(2002)}]{heger2002}
{Heger}, A., \& {Woosley}, S.~E. 2002, \apj, 567, 532.
  \eprint{arXiv:astro-ph/0107037}

\bibitem[{{Joggerst} \& {Whalen}(2011)}]{candace2010-2}
{Joggerst}, C.~C., \& {Whalen}, D.~J. 2011, \apj, 728, 129. \eprint{1010.4360}

\bibitem[{{Stacy} et~al.(2010){Stacy}, {Greif}, \& {Bromm}}]{stacy2010}
{Stacy}, A., {Greif}, T.~H., \& {Bromm}, V. 2010, \mnras, 403, 45.
  \eprint{0908.0712}

\bibitem[{{Timmes} \& {Swesty}(2000)}]{timmes2000}
{Timmes}, F.~X., \& {Swesty}, F.~D. 2000, \apjs, 126, 501

\bibitem[{{Turk} et~al.(2009){Turk}, {Abel}, \& {O'Shea}}]{turk2009}
{Turk}, M.~J., {Abel}, T., \& {O'Shea}, B. 2009, Science, 325, 601.
  \eprint{0907.2919}

\bibitem[{{Weaver} et~al.(1978){Weaver}, {Zimmerman}, \& {Woosley}}]{kepler}
{Weaver}, T.~A., {Zimmerman}, G.~B., \& {Woosley}, S.~E. 1978, \apj, 225, 1021

\bibitem[{{Whalen} et~al.(2008{\natexlab{a}}){Whalen}, {O'Shea}, {Smidt}, \&
  {Norman}}]{wet08a}
{Whalen}, D., {O'Shea}, B.~W., {Smidt}, J., \& {Norman}, M.~L.
  2008{\natexlab{a}}, \apj, 679, 925. \eprint{0708.1603}

\bibitem[{{Whalen} et~al.(2008{\natexlab{b}}){Whalen}, {van Veelen}, {O'Shea},
  \& {Norman}}]{wet08b}
{Whalen}, D., {van Veelen}, B., {O'Shea}, B.~W., \& {Norman}, M.~L.
  2008{\natexlab{b}}, \apj, 682, 49. \eprint{0801.3698}

\bibitem[{{Zhang} et~al.(2011){Zhang}, {Howell}, {Almgren}, {Burrows}, \&
  {Bell}}]{zhang2011}
{Zhang}, W., {Howell}, L.~H., {Almgren}, A.~S., {Burrows}, A., \& {Bell}, J.~B.
  2011, ApJ to appear

\end{thebibliography}

\end{document}